\documentclass{JINST}
\usepackage{amsmath,amsfonts,amssymb,fontenc,times,mathptmx,graphicx}
\title{Augmenting momentum resolution with well tuned probability distributions }
\author{Gregorio Landi$^a$\thanks{Corresponding
author.}~,   Giovanni E. Landi$^b$\\
\\
\llap{$^a$} Dipartimento di Fisica e Astronomia,
Universita' di Firenze\\
Largo E. Fermi 2 50125 Firenze Italy\\
and INFN, Sezione di Firenze,
Firenze,Italy\\
E-mail: \email{Gregorio.Landi@cern.ch}\\
\\
\llap{$^b$} E.T.T. S.a.g.l.,\\
Via Balestra 33,\\
6900 Lugano, Switzerland.\\}

\abstract{
The realistic probability distributions of a previous article are applied to the
reconstruction of tracks in constant magnetic field.
The complete forms and their schematic approximations produce
excellent momentum estimations, drastically better than standard fits.
A simplified derivation of one of our probability distributions is illustrated.
The momentum reconstructions are compared with standard fits (least squares) with two different position algorithms:
the $\eta_2$-algorithm and the two-strip center of gravity. The quality of our results are expressed as the increase of
the magnetic field and signal-to-noise ratio that overlap the standard fit reconstructions with ours best distributions.
The data and the simulations are tuned on the 
tracker of a running experiment and its double sided microstrip detectors, here each detector side is simulated to 
measure the magnetic bending.
To overlap with our best distributions, the magnetic
field must be increased by a factor 1.5 for the least squares based on the
$\eta_2$-algorithm and 1.8 for the two-strip center of gravity for the low noise side, and 1.8 and 2.0 for the high noise side.
The signal-to-noise ratio must be increased by 1.6 for the low noise side and 2.2 for the high noise side ($\eta_2$-algorithms).
The fits, built on the positioning with the center of gravity, are not modified by a reduction of the signal-to-noise ratio.}


\begin{document}

\section{Introduction }

Some properties of our well tuned probability density functions (PDFs) were described in ref.~\cite{landi05} and
were tested on fits of simulated straight tracks.
The limitation to straight tracks was mainly due to the
complexity of the method we applied for the first time. Hence, working with
two parameters, the debugging and testing of the application
can be followed on a surface. Furthermore, the data, elaborated in our approach, were collected
in a CERN test  beam~\cite{vannu} in the absence of a magnetic field.
The detectors we used, were few samples of double-sided silicon microstrip 
detector~\cite{aleph,L3} as those composing the PAMELA tracker~\cite{PAMELA}.
The results of ref.~\cite{landi05} showed a drastic improvement of
the fitted track parameters respect to the results of the least squares methods.   We
observed excellent reconstructions even in presence of very noisy hits, often called {\em outliers}, that generally
produce worse fits.
This achievement is almost natural for our non gaussian PDFs. In fact, an outlier hit is an
event that is incompatible with the gaussian distribution that is always assumed as the error distribution of the
hit positions.
Thus, being the least squares a method strictly
optimum for a gaussian PDF, it must be modified in a somewhat arbitrary way to handle these pathological hits.
On the contrary, our PDFs are essentially different from a
gaussian, and those non standard cases are allowed (with low probability) by the non-gaussian tails of the PDFs.

We have to recall that the perception of a rough handling of the hit properties is well present
in literature, and Kalman filter modifications are often studied to accept extended deviations from a pure gaussian model.
These extensions imply the use of linear
combinations of different gaussians~\cite{fru}, but in a small number to avoid the intractability of the equations.
The unknown parameters are optimized from an improvements of the fits.
In late sense, our schematic approximations, used to initialize the maximum likelihood search, could be reconnected to those
extensions. In fact, to speed the convergence, we calculate a different gaussian PDF (or more precisely an effective variance)
for each hit. 
Our comparisons take as references the least squares methods. The Gauss-Markov theorem states
their optimality among the linear methods, thus no other linear method can be better.

The confidence gained in ref.~\cite{landi05} with straight tracks allow us to deal with more complex tasks,
i.e. to reconstruct the tracks in a magnetic field and determine their momenta.
To face this extension with the minimum modifications of our previous work, we will  utilize the
simulated data and parameters
used in ref.~\cite{landi05}, adapting them to the geometry and average magnetic field of the PAMELA tracker.
In fact, the relatively low magnetic field introduces small modifications of the parameters
obtained in its absence. The average signal distribution of a minimum ionizing particle (MIP) is
slightly distorted by a Lorentz angle, that for the average magnetic field of the PAMELA tracker ( $0.43\, T$) is about
$0.7^\circ$. Around this incidence angle, the average signal distribution of a MIP is practically identical to that at
orthogonal incidence in the absence of a magnetic field, thus without further notice we will assume
a rotation of the detectors of their Lorentz angle. With these assumptions, we simulate high momentum tracks for
each side of the double sided detectors. In the PAMELA tracker, the low noise side of the detector
(the junction side) is used for momentum measurements. This side has an excellent resolution, a
strip each two is connected to the read out system and
the unconnected strip distributes the incident signal on the nearby ones optimizing the detector resolution.
The other side (the ohmic side) has the strips oriented perpendicularly to the junction side.
Each strip is connected to the readout system and has, by the complexity of construction, an higher noise
and a small, if any, signal spread to nearby strips. For those characteristics, this side responds in a way similar to the types
of microstrip arrays used in the large trackers of the CERN LHC~\cite{ALICE,ATLAS,CMSa}.
Thus, the simulations on this side, as bending side, can give a glimpse of our approach for other type of trackers even for
the (small angle stereo) double-sided detectors of the ALICE~\cite{ALICE} experiment.

\section{A compact derivation of the PDF}

As always, the MIP incidence of our geometry imposes the use of the minimum number of signal strips
to reduce the noise, hence only two signal strips will be used.
In ref.~\cite{landi05} we indicated the principal steps required to obtain the PDF for the center of gravity (COG) with two-strips, those steps followed the standard method described in the books about probability.
That procedure uses the essential tools of the calculus: integrals and derivatives.
In our cases, many pieces of integrals over geometrical domains must be
calculated to build the cumulative probability, and its derivative gives our final PDF. It is evident the length
and the complexity of this development.
Here we follow a very different approach (Quantum Mechanics style) that reaches identical
results in few steps.
The two strip COG (COG$_2$ in the following) is calculated with the origin in the center of the maximum-signal strip.
The strip signals are indicated with: $x_1$, $x_2$, and $x_3$, respectively the signal of the right strip,
central strip (with the maximum signal) and left strip. If $x_1>x_3$ the COG$_2$ is $x=x_1/(x_1+x_2)$,
if $x_3>x_1$ it is $x=-x_3/(x_3+x_2)$, other minor details will not be discussed here. Thus:
\begin{equation}\label{eq:equation_1}
\begin{aligned}
    P_{x_{g2}}(x)=&\int_{-\infty}^{+\infty}\mathrm{d}\,x_1\int_{-\infty}^{+\infty}\mathrm{d}\,x_2\int_{-\infty}^{+\infty}
    \mathrm{d}\,x_3\,P(x_1,x_2,x_3)\\
    &\,\Big[\theta(x_1-x_3)\delta(x-\frac{x_1}{x_1+x_2})+\theta(x_3-x_1)\delta(x+\frac{x_3}{x_3+x_2})\Big]
\end{aligned}
\end{equation}
where $P(x_1,x_2,x_3)$ is the probability to have the signals $x_1,x_2,x_3$ from the strips $1,2,3$. The signals
$x_i$ are at their final elaboration stage and ready to be used for
position reconstruction of the hit. The function
$\theta(x_j)$ is the Heaviside $\theta$-function: $\theta(x_j)=1$ for $x_j>0$, $\theta(x_j)=0$ for $x_j\leq 0$,
and $\delta(x)$ is the Dirac $\delta$-function.
It is immediate to verify the normalization of $P_{x_{g2}}(x)$ by direct integration.
Splitting the sum of eq.~\ref{eq:equation_1} in two independent
integrals and transforming the variables $x_1=\xi$, $x_1+x_2=z_1$ and $x_3=\beta$,
$x_3+x_2=z_2$, the jacobian of the transformation is one and the integrals in $z_1$ and $z_2$ can be
performed with the rule:
\begin{equation}\label{eq:equation_1a}
    \int_{-\infty}^{+\infty}\mathrm{d}\,z\, F(z-\nu)\,\delta(x\mp\frac{\nu}{z})=F(\frac{\pm\nu}{x}-\nu)\,\frac{|\nu|}{x^2}\, .
\end{equation}
Applying eq.~\ref{eq:equation_1a} to eq.~\ref{eq:equation_1} and using the limitations of the two
$\theta$-functions, eq.~\ref{eq:equation_1} becomes:
\begin{equation}\label{eq:equation_1b}
\begin{aligned}
    P_{x_{g2}}(x)=\frac{1}{x^2}\Big[&\int_{-\infty}^{+\infty}\mathrm{d}\,\xi\int_{-\infty}^\xi\,\mathrm{d}\beta\,
    P\big(\xi,\,\xi\frac{1-x}{x},\beta\big)\,|\xi|\,+\\
    &\int_{-\infty}^{+\infty}\mathrm{d}\,\beta\,\int_{-\infty}^\beta\,\mathrm{d}\xi\,P\big(\xi,\,
    \beta\frac{-1-x}{x},\beta\big)\,|\beta|\Big]\,.
\end{aligned}
\end{equation}
This form underlines very well the similarity with the Cauchy PDF; in the limit of $x\rightarrow\infty$ the $x$-part
of the $P$ arguments are $-1$ and $P_{x_{g2}}(x)\propto 1/x^2$ for large $x$.

\subsection{The probability $P_{x_{g2}}(x)$ for small $x$}

The probability $P(x_1,x_2,x_3)$ can handle a strict correlation among the arguments, we will release this strict correlation in a
weakest one: the mean values of the strip signals are correlated, but the fluctuations around the mean values are independent.
Thus the probability $P(x_1,x_2,x_3)$ becomes the product of three functions $\{P_i(x_i),i=1,2,3\}$.
Each $P_i(x_i)$ is assumed to be a gaussian PDF with mean
values $a_i$ and standard deviation $\sigma_i$. To simplify, the constants $a_i,i=1,2,3$ are the noiseless signals
released by a MIP with impact point $\varepsilon$:
\begin{equation}\label{eq:equation_1d}
    P_i(x_i)=\exp\big[-\frac{(x_i-a_i)^2}{2\sigma_i^2}\big]\frac{1}{\sqrt{2\pi}\sigma_i}.
\end{equation}
Even with the gaussian functions, the integrals of eq.~\ref{eq:equation_1b} have no analytical expressions and effective
approximations must be constructed. We will not report our final forms that are very long, instead we will illustrate a limiting
case which gives a simple approximation and eliminates a disturbing singularity for the numerical integrations.
It easy to show that for $|x|\rightarrow 0$, $x^{-1}P_2\big(\xi(1-x)/x\big)$ and
$x^{-1}P_2\big(\beta(-1-x)/x\big)$ converge to two Dirac $\delta$-functions. Hence, for small $|x|$,
the integrals of eq.~\ref{eq:equation_1b} can be expressed as:
\begin{equation}\label{eq:equation_1c}
\begin{aligned}
    P_{x_{g2}}(x)=&\frac{|a_2|}{\sqrt{2\pi}}\Big\{
    \frac{\exp\big[-(x-\frac{a_1}{a_1+a_2})^2\frac{(a_1+a_2)^2}{2\sigma_1^2(1-x)^2}\big]
    \big[1-\mathrm{erf}\big((\frac{a_3}{a_2+a_3}-x)\frac{a_2+a_3}{\sqrt{2}\sigma_3(1-x)}\big)\big]}{2\sigma_1(1-x)^2}+\\
    &\frac{\exp\big[-(x+\frac{a_3}{a_3+a_2})^2\frac{(a_3+a_2)^2}{2\sigma_3^2(1+x)^2}\big]
    \big[1-\mathrm{erf}\big((\frac{a_1}{a_2+a_1}+x)\frac{a_2+a_1}{\sqrt{2}\sigma_1(1+x)}\big)\big]}{2\sigma_3(1+x)^2}\Big\}\,.
\end{aligned}
\end{equation}
Equation~\ref{eq:equation_1c} is correct only for $|x|\rightarrow 0$, but it is useful beyond this limit and contains many ingredients
of more complex expressions. An example can be seen in fig.2 of ref.~\cite{landi05}, the essential elements
are the two bumps (gaussian-like)  centered in the possible noiseless two-strip
COG: $a_1/(a_1+a_2)$ and $-a_3/(a_3+a_2)$.
Their effective standard deviations are modulated by the signal-to-noise ratio $\sigma_1/(a_2+a_1)$ and $\sigma_3/(a_2+a_3)$
and by the $1\pm x$ factor.
More complete expressions contain terms very similar to Cauchy PDFs  that are $\propto 1/x^2$ for large $x$.
The dimensions of the constants $a_j$ must be those of the $\sigma_j$, for both of them we take directly the
ADC counts. The $x$-variable (the COG$_2$) is a pure number expressed as a fraction of the strip size, or more precisely, the strip
size is the scale of lengths.

The form of the COG$_2$ inserted in eq.~\ref{eq:equation_1} synthesizes well our positioning
algorithm: each strip signal around
the strip with the maximum signal is used for position reconstruction.
The strategy of cluster detection is
supposed to be optimized to suppress the false hits and its results are used only for the
maximum-signal-strip selection, and the signals of the two lateral strips are used in any case
even for small positive or not too negative values. It must be underlined that the tiny amount of
information is relevant for positioning or to maintain the noise distribution.
In fact, the track reconstruction algorithms reach their optimum with
realistic probability distributions, and any unaccounted distortion induces a loss of resolution.

For construction, $P_{x_{g2}}(x)$ gives to the probability of an $x$ value with constant $\{a_j\}$.
For the track fitting this simple probability is useless, we need a PDF that, for any $x$-value, gives the probabilities
of the impact points $\{\varepsilon\}$. To insert this functional dependence on $P_{x_{g2}}(x)$, we must know the average variations
of the energies $\{a_1,a_2,a_3\}$ with $\varepsilon$. For this task, it is essential
a theorem illustrated in ref.~\cite{landi05} and the details of refs.~\cite{landi01,landi03,landi04}, this theorem  allows
the extraction of these functional dependencies from the data. Once extended in $\varepsilon$, the PDF can be rewritten as:
\begin{equation}\label{eq:equation_2}
    P_{x_{g2}}(x,E_t,\varepsilon)=\frac{F(a_1(\varepsilon),a_2(\varepsilon),a_3(\varepsilon),E_t,\sigma_1,\sigma_2,\sigma_3,x)}{x^2}\,.
\end{equation}
Where $F(a_1(\varepsilon),a_2(\varepsilon),a_3(\varepsilon),E_t,\sigma_1,\sigma_2,\sigma_3,x)$ is
the term in square brackets
of eq.~\ref{eq:equation_1b}. The functions $a_1(\varepsilon),a_2(\varepsilon),a_3(\varepsilon)$ are the
average noiseless fractions of signal collected by each strip and $E_t$ is the total signal
of the three strips: the central one with the maximum signal and the two lateral. We would need
the noiseless $E_t$, but the measured one is the only possibility.
The extraction of the functions $\{a_i(\varepsilon)\}$ from
real data of ref.~\cite{landi05} necessitates further refinements and numerical recipes.
But, once defined and checked their results, the procedure can be
standardized. In any case, slight variations of the $\{a_i(\varepsilon)\}$ around the best one give almost
identical track parameter distributions, thus their selection is very important but non critical.
The functions $\{a_i(\varepsilon)\}$ have a definition similar to the "templates" of ref.~\cite{CMS}.
The parameters $\sigma_1,\sigma_2,\sigma_3$
are the noise (eq.~\ref{eq:equation_1d}) of the three strips considered. 
Our PDF can easily accomodate strips with different $\sigma$, but,
in the simulations, we will use a single $\sigma$ for all the strips of same detector side and these parameters will not be
reported in the future expressions. Equation~\ref{eq:equation_2} is normalized as function of $x$ but
is not normalized in $\varepsilon$, due to its irrelevance for the fit, this normalization will be neglected here.
A set of normalized PDFs from eq.~\ref{eq:equation_2} are illustrated in ref.~\cite{landi05} as $\varepsilon$ functions.

\subsection{Track definition}

Given our exploration of a new approach, our track definition must be the easiest allowed.
The tracks are circles with a large radius to simulate the high momentum MIPs
where the multiple scattering is negligible. The relation of the track parameters to the momenta is $p=0.3\, \mathrm{B}\, R$,
the form, adapted to our needs, from ref.~\cite{particle_group} where $p$ is in $\mathrm{GeV/c}$, $\mathrm{B}$ in Tesla and $R$, the track radius, in meters.\newline
The tracker model is formed by six parallel equidistant ($89\, mm$) detector layers as for the PAMELA tracker, with constant
magnetic field of $0.43\,T$ perpendicular to the track plane ($\xi\,,z$). The center of the tracker has coordinates
$\xi=0\,,z=0$, the $\xi$ axis is parallel to the layer planes and the $z$ axis is perpendicular.
The tracks are circles with center in $\xi=-R$ and $z=0$, and the magnetic field is parallel to the analyzing strips.
To simplify the geometry, the overall small rotation of the Lorentz angle ($0.7^\circ$)  is neglected now, but it will
be introduced in the following.
The simulated hits are generated (ref.~\cite{landi05}) with a uniform random distribution on a strip,
they are collected in groups of six to produce the tracks. For the least squares, the exact impact position $\varepsilon$ of each hit is subtracted from
its reconstructed $\eta_2(x_{g2})$ position (as defined in refs.~\cite{landi05,landi03,belau}) and it is added the value of the
fiducial track for the corresponding detector layer. In this way each group of six hits
defines a track with our geometry and the error distribution of $\eta_2$-algorithm. Identically for the
COG$_2$ positioning algorithm.
This hit collection
simulates a set of tracks populating a large portion
of the tracker system with slightly non parallel strips on different layers (as it is always  in real detectors).
All these tracks are possible realizations of our model track.
At our high momenta the track bending is very small and the track sagitta is smaller than the strip width.
Thus, the bunch of tracks has a transversal section around a strip size, on this width we have to
consider the Lorentz angle but its effect is clearly negligible.
Our preferred position reconstruction is the $\eta_2$ algorithm as in ref.~\cite{landi05}  because it
gives parameter distributions better than those obtained with the simplest COG$_2$
positions, but even the results for the COG$_2$ will be reported in the following.\newline
In the $\{{\xi},{z}\}$-plane, the circular tracks are approximated, as usual,  with parabolas linear in the
track parameters:
\begin{equation}\label{eq:equation_3}
\begin{aligned}
    &\xi=\beta+\gamma {z}-\alpha {z}^2=\varphi(z)\,\\
    &\xi_n=\beta_n+\gamma_n {z}-\alpha_n {z}^2=\varphi_n(z)\,.\\
\end{aligned}
\end{equation}
The first line of eq.~\ref{eq:equation_3} is
the model track, the second line is the fitted one.
The circular track is the osculating circle of the parabola $\varphi(z)$,
at our high momenta and tracker size the differences are negligible.
Our model track has $\gamma=0,\,\beta=0$, and $1/\alpha$ is proportional to the
track momentum. Due to the noise, the reconstructed track has equation $\varphi_n(z)$ and
the fitted parameters $\{\alpha_n,\beta_n,\gamma_n\}$ are distributed around the model values $\alpha,\beta,\gamma$.
For their non gaussian forms, our PDFs must be used with the non-linear search of the likelihood
maxima, but, as always, the search will be transformed in a minimization. The momentum
and the other parameters of the track $n$ are obtained minimizing, respect to $\alpha_n$, $\beta_n$ and $\gamma_n$ the function
$L(\alpha_n,\beta_n,\gamma_n)$ defined as the negative logarithm of the likelihood with the
PDFs of eq.~\ref{eq:equation_2}:
\begin{equation}\label{eq:equation_4}
\begin{aligned}
    &L(\alpha_n,\beta_n,\gamma_n)=-\sum_{j=6n+1}^{6n+6}\,\ln[P_{x_{g2}}(x(j),E_t(j),\psi_j(\alpha_n, \beta_n,\gamma_n)]\\
    &\psi_j(\alpha_n,\beta_n,\gamma_n)=\varepsilon(j)-\varphi(z_j)+\varphi_n(z_j)\,.
\end{aligned}
\end{equation}
The parameters $x(j)$, $E_t(j)$ (introduced in eq.~\ref{eq:equation_2}) are respectively:
the COG$_2(j)$ position, the sum of signal in the three strips for the hit $j$ and the track $n$,
${z}_j$ is the position of the 
detector plane $j$ of the $n$th track.
The $\varepsilon$-dependence in the $\{a_i(\varepsilon)\}$ is modified to $\psi_j(\alpha_n,\beta_n,\gamma_n)$,
to place the impact points on the track. In real data, $\varepsilon(j)-\varphi(\alpha,\beta,\gamma,z_j)$ is
absent (and unknown) but the data are supposed to be on a track. We can easily use  a non linear form for the
function $\psi_j(\alpha_n,\beta_n,\gamma_n)$, but in this case is of scarce meaning. In more complex cases, 
non linearities of various origin can be easily implemented.

We will reserve the definition of maximum likelihood evaluation (MLE) to the results of eq.~\ref{eq:equation_4} even
if the least squares method can be derived by a maximum likelihood.
The minimum search routine for the MLE  is initialized as in ref.~\cite{landi05}. The initial track parameters are
given by a weighted least squares with weights given by an effective variance ($\sigma_{eff}(i)^2$)
for each hit $(i)$ and $\eta_2(i)$ as hit position. The $\sigma_{eff}(i)^2$ is obtained from our PDF, but, for the form of
eq.~\ref{eq:equation_2}, the variance is an hill defined parameter even in
$\varepsilon$, and this hill definition must be eliminated with cuts on the integration ranges that
suppress the PDF tails. We use two sizes of cuts,
one for the low noise side and one for the high noise side of our double side detector.
The cuts are optimized to obtain gaussian distributions, with standard deviation $\sigma_{eff}(i)$, reproducing
our PDF for a good hit.
For a set of hits (good hits), eq.~\ref{eq:equation_2} has the form of a narrow high peak
and a gaussian (centered in
$\eta_2(i)$) with standard deviation $\sigma_{eff}(i)$ is built to reproduce well the $P_{x_{g2}}(x(i),E_t(i),\varepsilon)$ on
few of them.
These gaussian approximations look good in linear plots, the logarithmic plots show  marked differences even
in these happy cases, the tails are non-gaussian. The cuts,
so defined, are used for the $\sigma_{eff}(i)$ extraction in all the other hits, even where $P_{x_{g2}}(x(i),E_t(i),\varepsilon)$ is
poorly reproduced by a gaussian.

The $\{\alpha_n,\beta_n,\gamma_n\}$ given by the weighted least squares are almost always near those given by the minimization of eq.~\ref{eq:equation_4}, thus rendering less heavy the minimum search to the MATLAB~\cite{MATLAB}
{$fminsearch$} routine.
The closeness of these approximations to the MLE supports the
non criticality of the extraction of the
functions $\{a_i(\varepsilon)\}$. In fact, the approximate gaussian distributions
are often very different from the $P_{x_{g2}}(x(i),E_t(i),\varepsilon) $, but the realistic pieces of information
about the hits are sufficient to produce near optimal results. When the tails of the PDFs are important the MLE results are better.
The strong variations of the set $\{\sigma_{eff}\}$ on the strip are illustrated in figs. 6 and 11 of ref.~\cite{landi05}, and the patterns of those variations support a complex interplay among the physical properties of
the strips and the reconstruction algorithm. The time-consuming extraction of $\sigma_{eff}(i)$ (as indicated in ref.~\cite{landi05})
can be reduced building the surface $\sigma_{eff}(x,E_t)$ and calculating $\sigma_{eff}(i)$ with an interpolation.

\section{Low noise, high resolution, floating strip side}

The floating strip side is the best of the two sides of this type of strip detector. It is just this
side that measures the track bending in the PAMELA magnetic spectrometer.
In the test beam~\cite{vannu}, the noise of the "average" strips is well reproduced by a gaussian with a standard
deviation of 4 ADC counts, and the PDF for the sum of three strip signals has its maximum at 142 ADC counts with the 
most probable signal-to-noise ratio of $35.5$ for a three strip cluster  ($SNR(n)=\sum_{i=1}^3 x_i(n)/\sigma_i$).
The functions $a_j(\varepsilon)$ are those of
ref.~\cite{landi05} for this strip type.
In the simulations we will use a high momentum of $350 \, GeV/c$. For this momentum and similar geometry,
we have a report with some histograms of a CERN test beam of the PAMELA tracker before its installation in the satellite.
The tracks were reconstructed with the $\eta_2$ algorithm, and this allows a positive check of our simulations.

At $350 \, GeV/c$, the hits of the tracks are largely contained
\begin{figure}[ht]
\begin{center}
\includegraphics[scale=0.5]{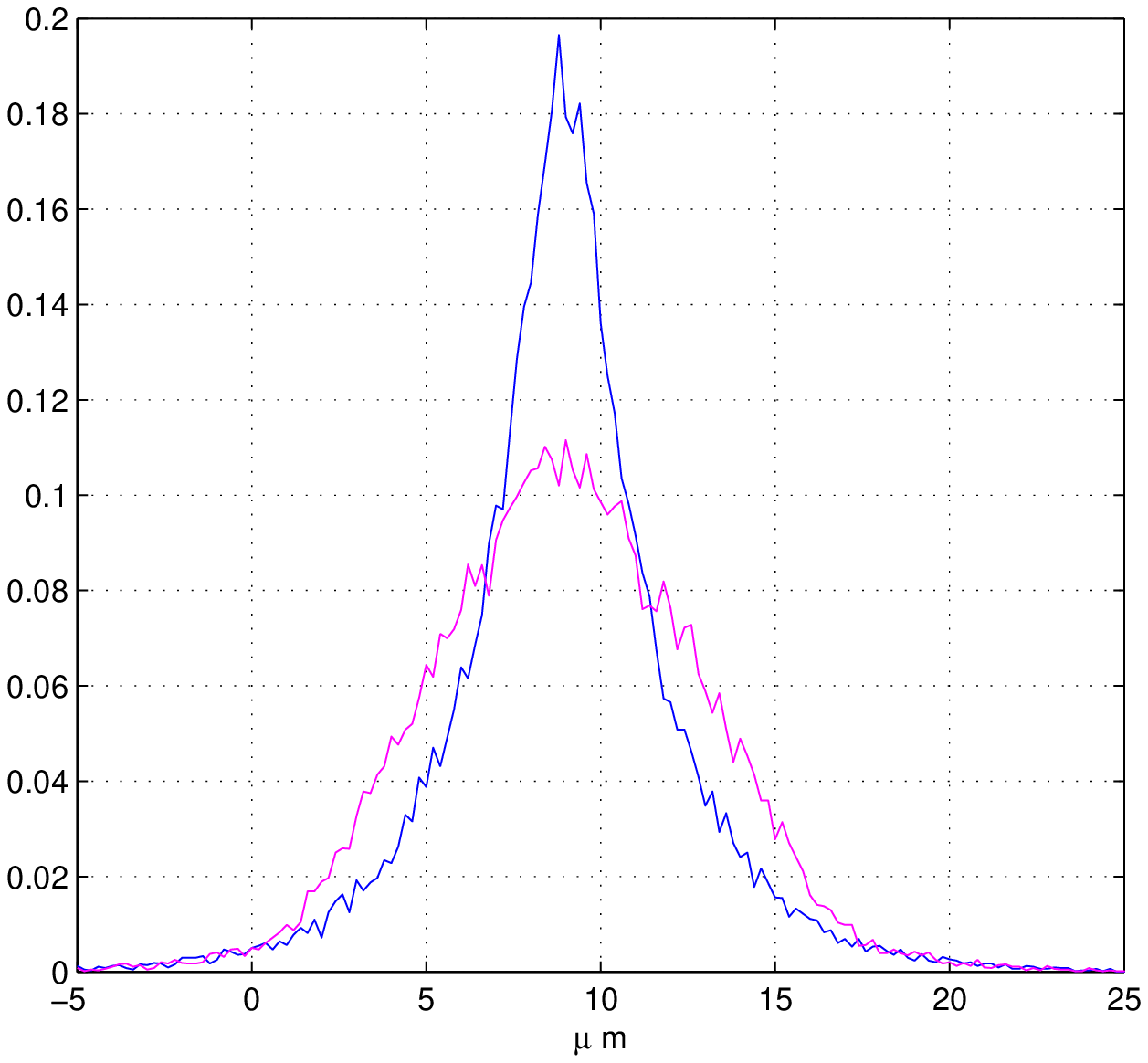}
\includegraphics[scale=0.5]{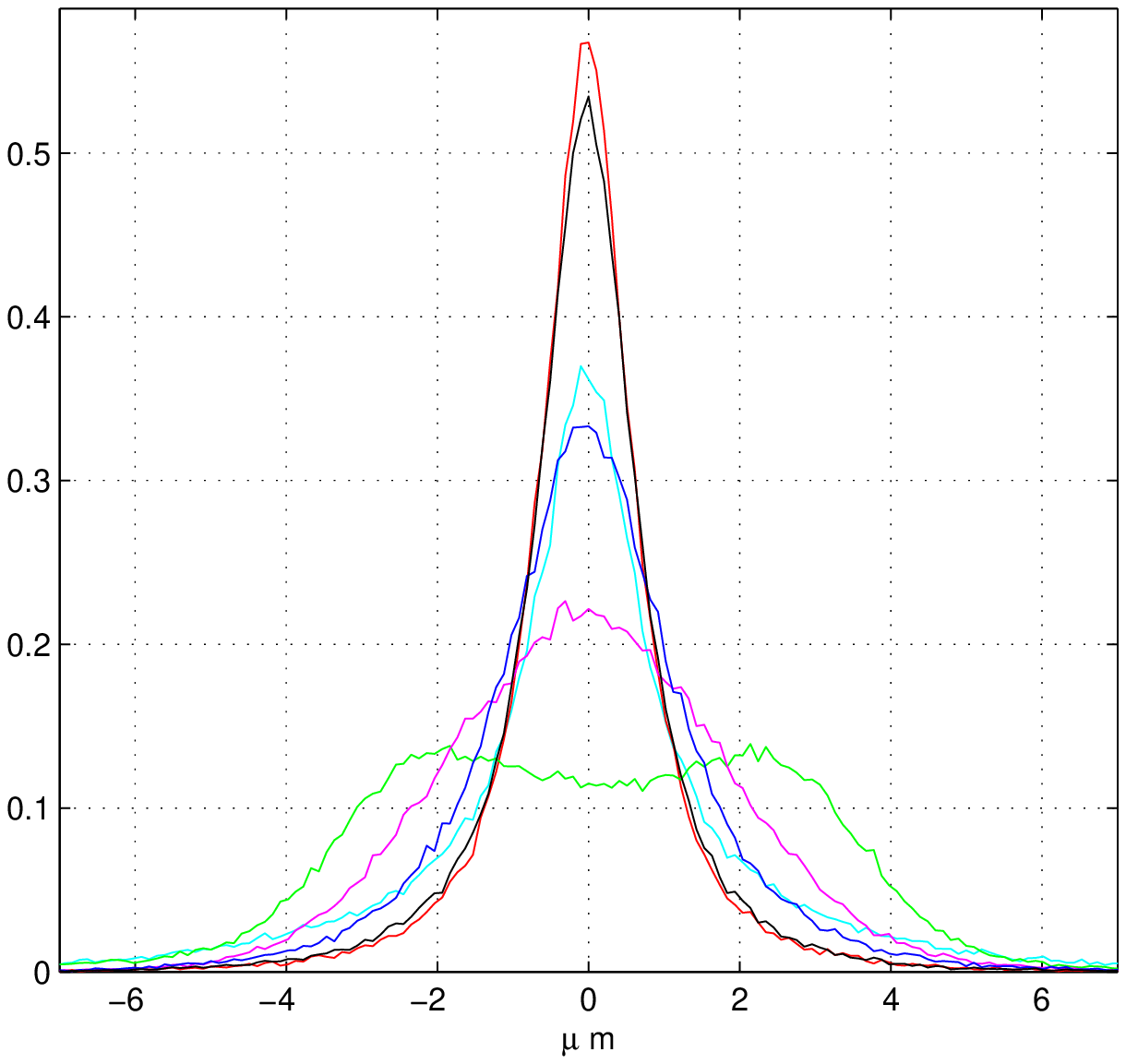}
\caption{\em Left plot. Blue line: distribution of the differences of hit $\#\,3$ minus  hit $\#\,1$ of a track for the $\eta_2$-algorithm; magenta line: the same for the COG$_2$. Right plot. True residuals of the reconstructed tracks: from MLE (red), from $\sigma_{eff}(i)$
(black), from $\eta_2$ (blue), from COG$_2$ (magenta), and the position errors of the $\eta_2$ (cyan)
and COG$_2$ (green).
}\label{fig:figure01}
\end{center}
\end{figure}
in a strip width ($51\, \mu m$), the left side of fig.~\ref{fig:figure01} shows the distributions of the
differences between two hits on a track
(the $\#\,3$ minus the $\#\,1$) for two different hit reconstruction algorithm.
They peak around 8 $\mu m$, but, as any COG reconstruction algorithm, the COG$_2$
algorithm produces a wider distribution due to a systematic error of refs.~\cite{landi01}.
The $\eta_2$-algorithm is  built to be free of this systematic error, here we corrected even the
asymmetry errors discussed in ref.~\cite{landi03,landi04}. Even if they are now very small,
the clear symmetry of the plots  is the product of this accuracy. For the absence of the
COG systematic error, the PDFs of the track parameters
given by the $\eta_2$ algorithm are  better than those given by the COG$_2$.
The results of our MLE are compared with those obtained by three different least squares.
Having to plot the results of four type of reconstructions we will use the following color convention:
\begin{itemize}
  \item {\bf{red lines}} refer to our MLE  (eq.~\ref{eq:equation_4}),
  \item {\bf black lines} are the weighted least squares with weight $1/\sigma_{eff}(i)^{2}$ and $\eta_2$ position,
  \item {\bf blue lines} for the least squares with the $\eta_2$ position algorithm
  \item {\bf magenta lines} for the least squares with the COG$_2$ position algorithm.
\end{itemize}
In the right plot
of fig.~\ref{fig:figure01} the distributions (or more precisely the histogram values divided by the number of entries and the step size) of the differences of the fitted positions respect to the exact ones are reported with the above color convention.
In the following we will call these differences as {\em true residuals} (allowed only in simulations) to distinguish from the
{\em residuals} generally defined as the
differences of the fitted positions from the reconstructed positions.
For further tests, two other lines are added, the cyan and green lines are the error distributions of the $\eta_2$ and COG$_2$
hit reconstruction algorithms. The PDFs of the residuals are not reported, but, those for the COG$_2$ and $\eta_2$ are
almost identical to the reported PDFs of the true residual. The residuals for other two approaches are very different
from the true residuals of fig.~\ref{fig:figure01}, they have very high peaks around zero. Our PDF of eq.~\ref{eq:equation_2}
and $\sigma_{eff}(i)$ allow the recognition of the good hits and the fit optimization selects to pass near to them, giving
an high frequency of small residuals.

The general expectancy of an improvement of the position reconstruction by redundant data is evidently disappointed
for the $\eta_2$ least squares: the cyan distribution is higher than the blue one. This mismatch is very similar to the case of a least squares on data from a Cauchy distribution~\cite{landi06}.
Instead, the  true residuals of COG$_2$ least squares look much better than their error distribution,
that is very large. The least squares looks able to round the error distribution.
But this effort consumes all its power, in fact if the
statistical noise is suppressed, the COG systematic error does not allow any modification of the true residuals.
On the contrary,  the true residual PDF of the $\eta_2$ least squares grows toward a Dirac $\delta$ function in the absence of
the statistical noise. We have to signal a light inconsistency among the histograms reported here and those of ref.~\cite{landi05}.
There we worked always with the strip pitch as scale for the length (as defined for the probability of eq.~\ref{eq:equation_1c}),
in the plots the horizontal scale was turned to the appropriate scales ($\mu m$). Here, the conversion to the correct dimensions
was performed before building the histograms, thus the vertical scales turn out different respect to those of ref.~\cite{landi05}.

\begin{figure}[h]
\begin{center}
\includegraphics[scale=0.53]{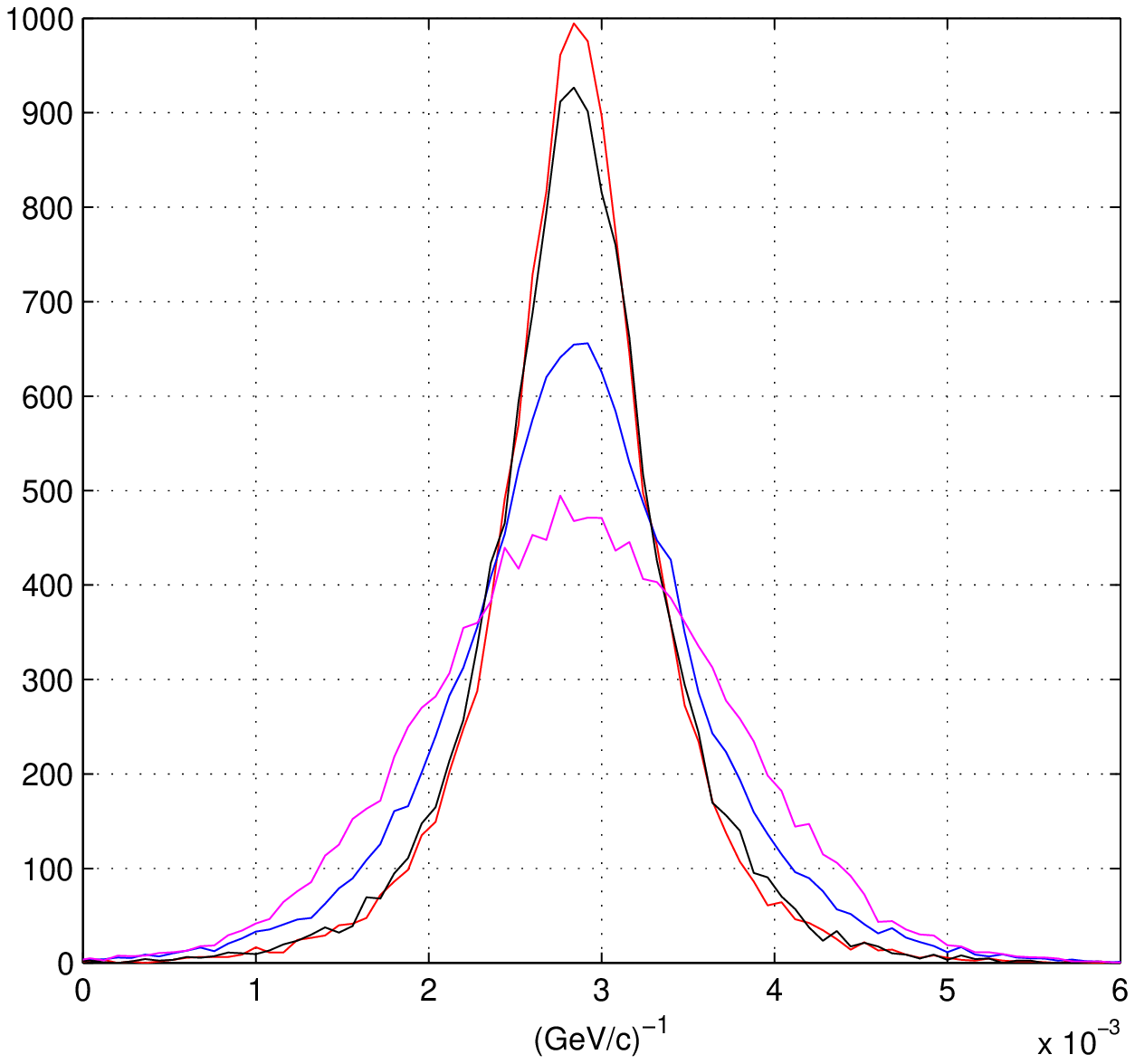}
\includegraphics[scale=0.53]{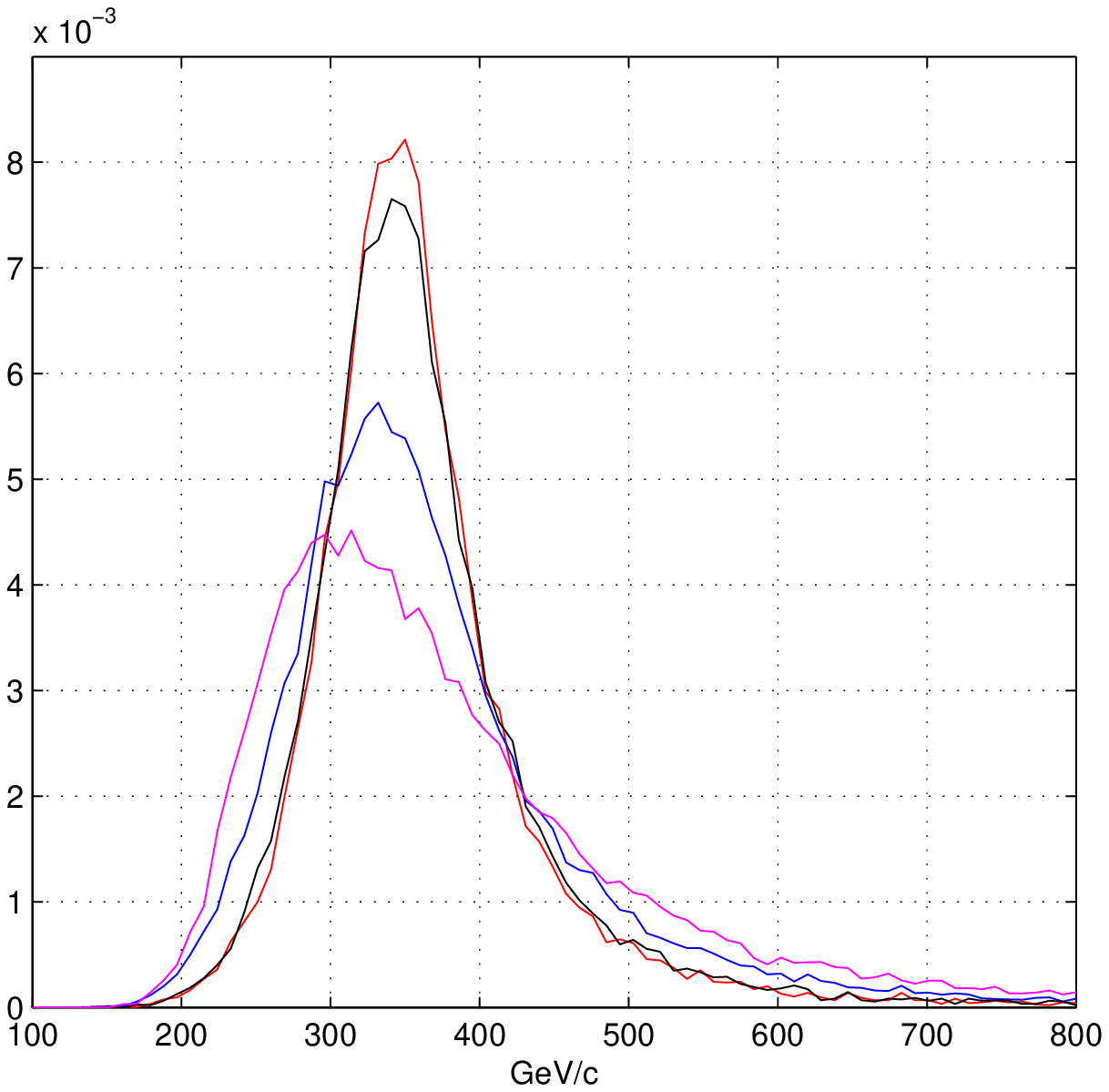}
\caption{\em Left plot. Distributions of the curvature in $(GeV/c)^{-1}$ (the $\alpha$ parameters) for the four types of fit. Right plot: distributions of the reconstructed momenta.
}\label{fig:figure02}
\end{center}
\end{figure}
As illustrated in fig.~\ref{fig:figure02}, the MLEs  give the best results for the momentum reconstruction,
and the weighted least squares, are very near to them. The fits of the standard
least squares with $\eta_2$
or COG$_2$ positioning algorithms show a drastic decrease in resolution. The use of the simple
COG$_2$ algorithm is the worst one.
Often the distributions of the left side of fig.~\ref{fig:figure02} are reported as resolution of the momentum reconstruction,
the k-value of ref.~\cite{particle_group}. For the $\eta_2$ and COG$_2$ least squares, the plots of the momentum distributions have appreciable shifts of the maxima
(most probable value) respect to the fiducial value of 350 $GeV/c$, the shifts are negligible for the other two
fits. These shifts are mathematical consequences of the change of variable from curvature to momentum.

\subsection{Other track parameters}

The complete track reconstruction must consider even the other two parameters of a track, the $\beta_n$
and $\gamma_n$. Their fits give very similar distributions to those plotted in ref.~\cite{landi05}.
The maxima of the distributions are now a little lower, in particular for the $\beta$ parameter. This is not unexpected,
in ref.~\cite{landi05} we had 3 degree of freedom for two parameters, here we have 3 degree of freedom for three
parameters, an effective reduction  of the redundancy that has a slight effect on the results.

\section{High noise, low resolution, normal strip side}

The other side of the double sided silicon microstrip detector has very different properties respect to the floating strip
side (the junction side). We will not recall the special treatments required to transform a ohmic side
in strip detector  and all the other particular setups necessary to its functioning. From the point of view of their data,
this side produces data very similar to a normal strip with a gaussian noise of 8 ADC counts,
twice of the other side (most probable signal-to-noise ratio $SNR(n)=18.2$). The absence of the floating strips gives to the histograms of the COG$_2$ the normal aspect with a high central density and a drop around $x_{g2}=0$.
No additional rises around
$x_{g2}=\pm1/2$ are present, they are typical of the charge spreads given by the floating strips. This absence reduces the
efficiency of the positioning algorithms that gain substantially from the charge spread. Now, the functions
$\{a_i(\varepsilon)\}$ of ref.~\cite{landi05} are similar to those of an interval function with a weak rounding to the borders,
this rounding is mainly due to the convolution of the strip response with
the charge spread produced by the drift in the collecting field. If a residual capacitive coupling is present, it
is very small. In any case, it is just this rounding that renders very good the resolution of the hits on the strip borders
with a small $\sigma_{eff}(i)$. Due to the differences respect to the other side, we have to implement the simulations
with a lower momentum of $150\, GeV/c$.

\begin{figure}[ht]
\begin{center}
\includegraphics[scale=0.53]{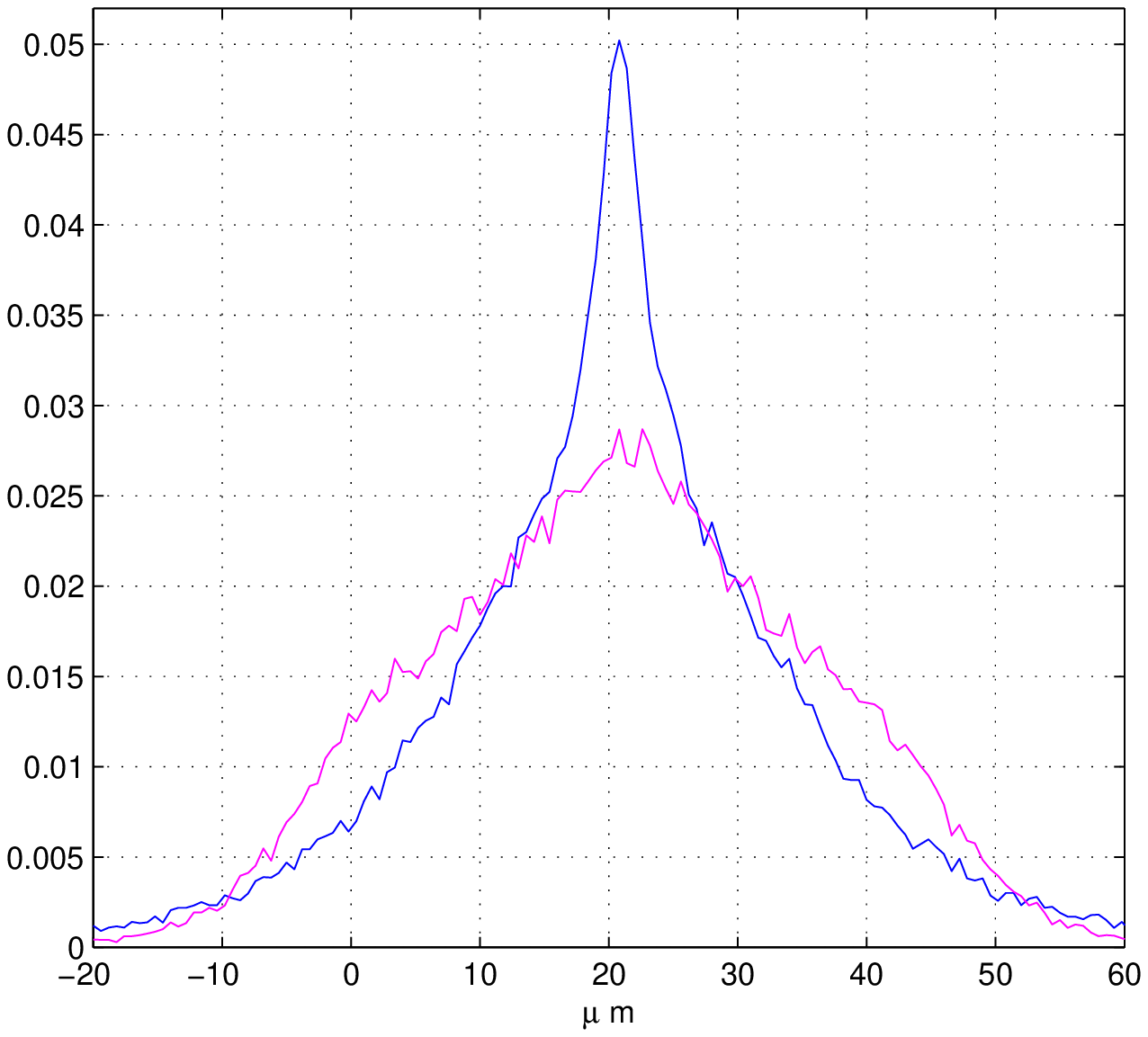}
\includegraphics[scale=0.53]{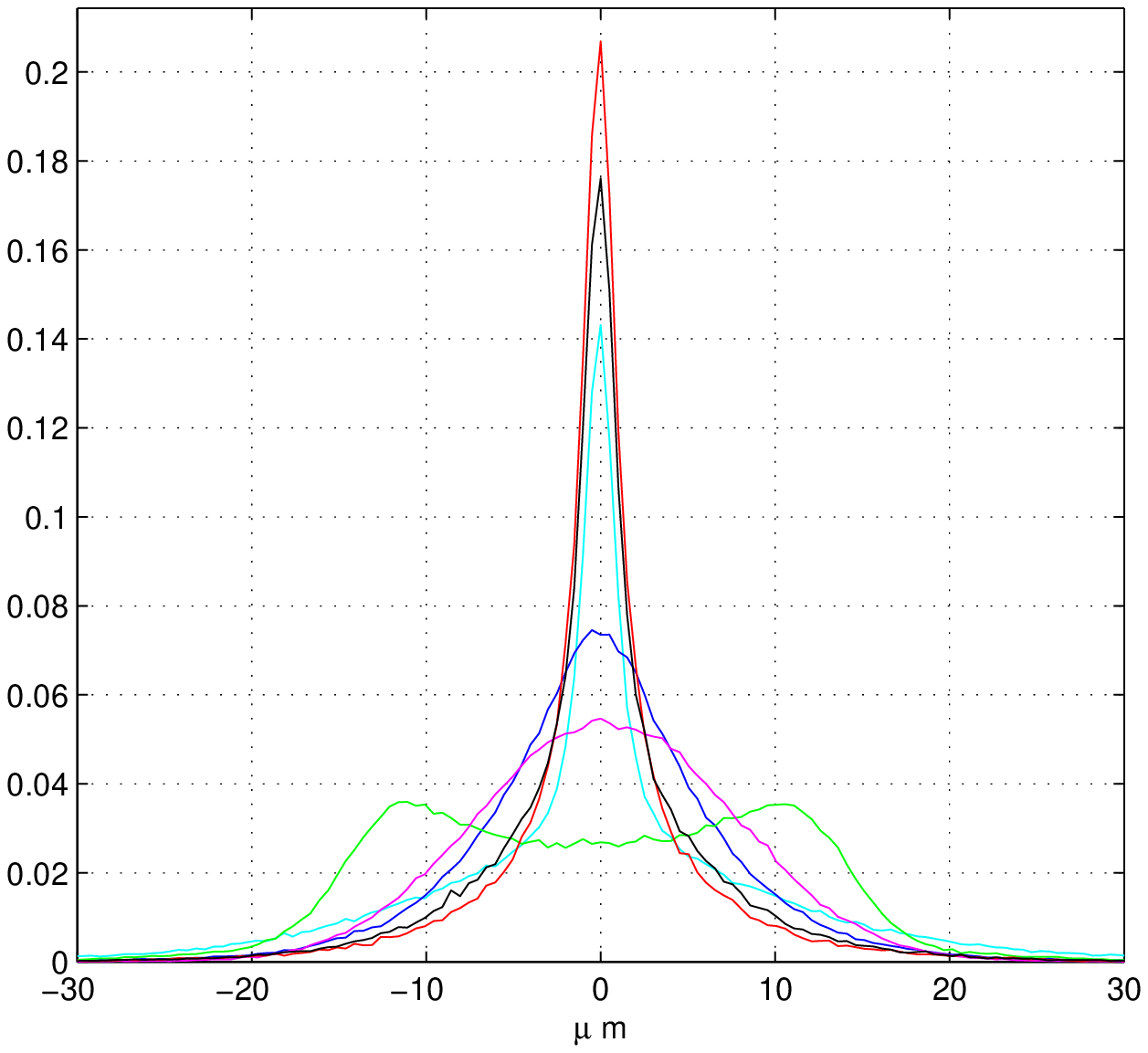}
\caption{\em Left plot. Blue line: distribution of the differences of hit $\#\,3$ minus  hit $\#\,1$ of a track for the $\eta_2$-algorithm, magenta line: the same for the COG$_2$. Right plot. True residuals of the reconstructed tracks. Color cades
defined above for the fits, the cyan and green PDFs are the positioning errors of the $\eta_2$
and COG$_2$ algorithms.
}\label{fig:figure03}
\end{center}
\end{figure}

\begin{figure}[h]
\begin{center}
\includegraphics[scale=0.53]{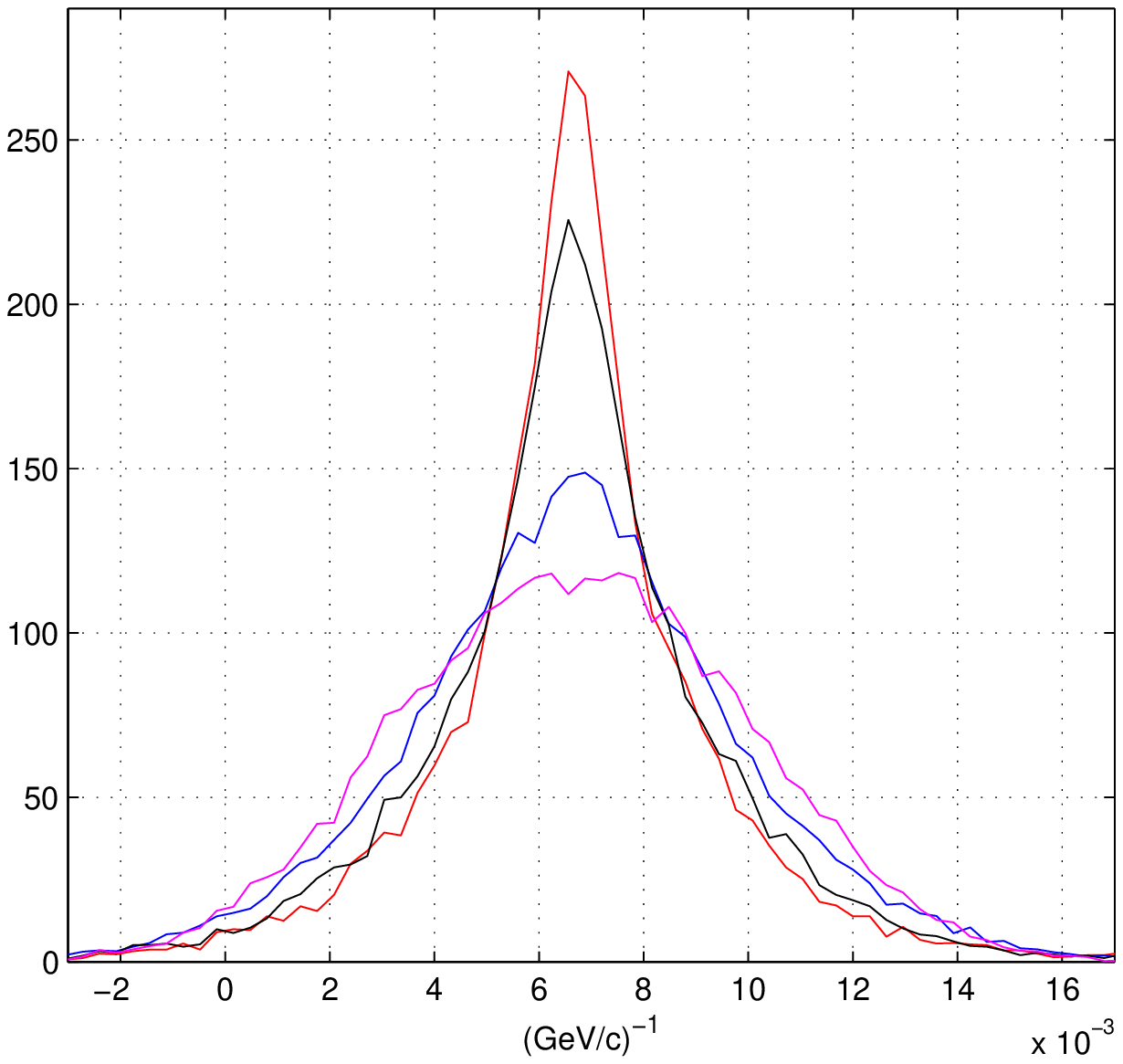}
\includegraphics[scale=0.53]{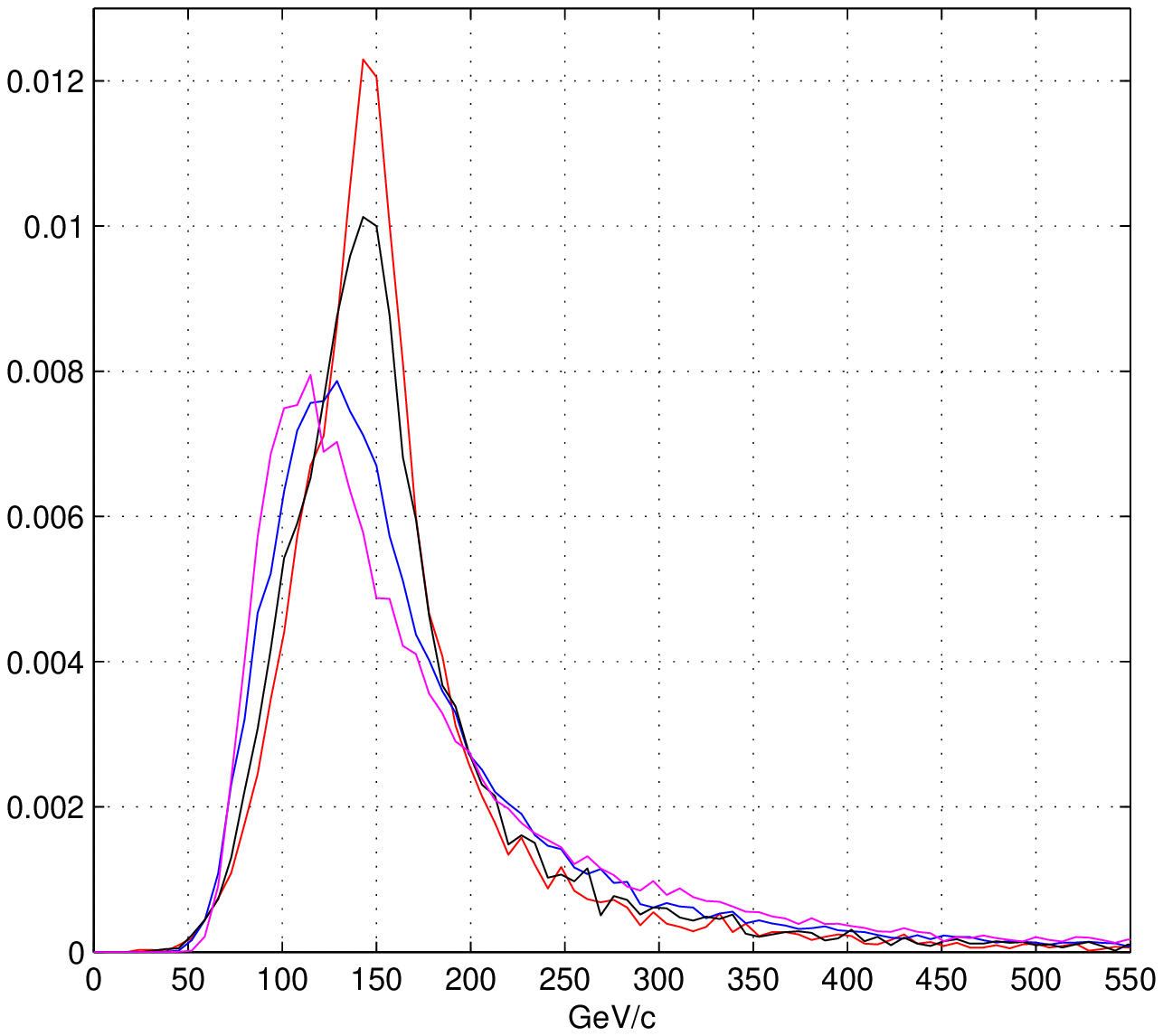}
\caption{\em Left plot. Distributions of the curvature in $(GeV/c)^{-1}$ (the $\alpha$ parameters) for the four
different fits. Right plot: distributions of the reconstructed momenta.
}\label{fig:figure04}
\end{center}
\end{figure}

The left side of fig.~\ref{fig:figure03} shows the distributions of the hit differences ($\#\,3$ minus $\#\,1$) to test the containment of the track on a strip width, the largest noise has evident effects on their forms. As usual the COG$_2$ distribution is larger than
the $\eta_2$ distribution. The true residuals of the fits are reported in the right side of fig.~\ref{fig:figure03} with the cyan and green lines for the error distributions of the $\eta_2$
and COG$_2$ hit reconstruction algorithms. The other color codes are the standard ones. Even here the data
redundancy for the $\eta_2$-fit does not improve the position reconstruction. The $\eta_2$ hit error
distribution is drastically better than that of the true residuals for the fit. The COG$_2$ position
error distribution has two maxima due to the
systematic error that is positive for the first half of the strip and negative in the second half,
this difference of sign combines with the noise random error to round the two maxima. The fit,
based on the COG$_2$ positions, has a good probability to partially average these sign differences
and produce a single wide maximum. The momentum distributions given by the four fits are reported
in fig.~\ref{fig:figure04}.

The results of our MLE and the weighted least squares are drastically better than the standard least squares.
Now the shift of the maxima for the momentum distributions of two least squares are more evident than in
fig.~\ref{fig:figure02}.

\section{Discussion}

A meaningful comparison among the different fits is somewhat difficult. Often the standard deviation of the PDF
is extracted interpolating a gaussian
function on non gaussian distributions, or two (or more) gaussian PDF are fitted hoping to describe two (or more) independent
gaussian processes. In ref.~\cite{landi05} we used the full width at half maximum, but, apart from its complex extraction,
it does not characterize very well the
distributions. In any case those differences are not of easy interpretation. Here we will use
two very precious resources contributing to the momentum measurements: the magnetic field and the signal-to-noise ratio.
The simulated magnetic field intensity and the signal-to-noise ratio are increased in the $\eta_2$ and COG$_2$ least squares
reconstructions to reach our best distributions (the red lines).

\subsection{Increasing the magnetic field}

\begin{figure}[h]
\begin{center}
\includegraphics[scale=0.45]{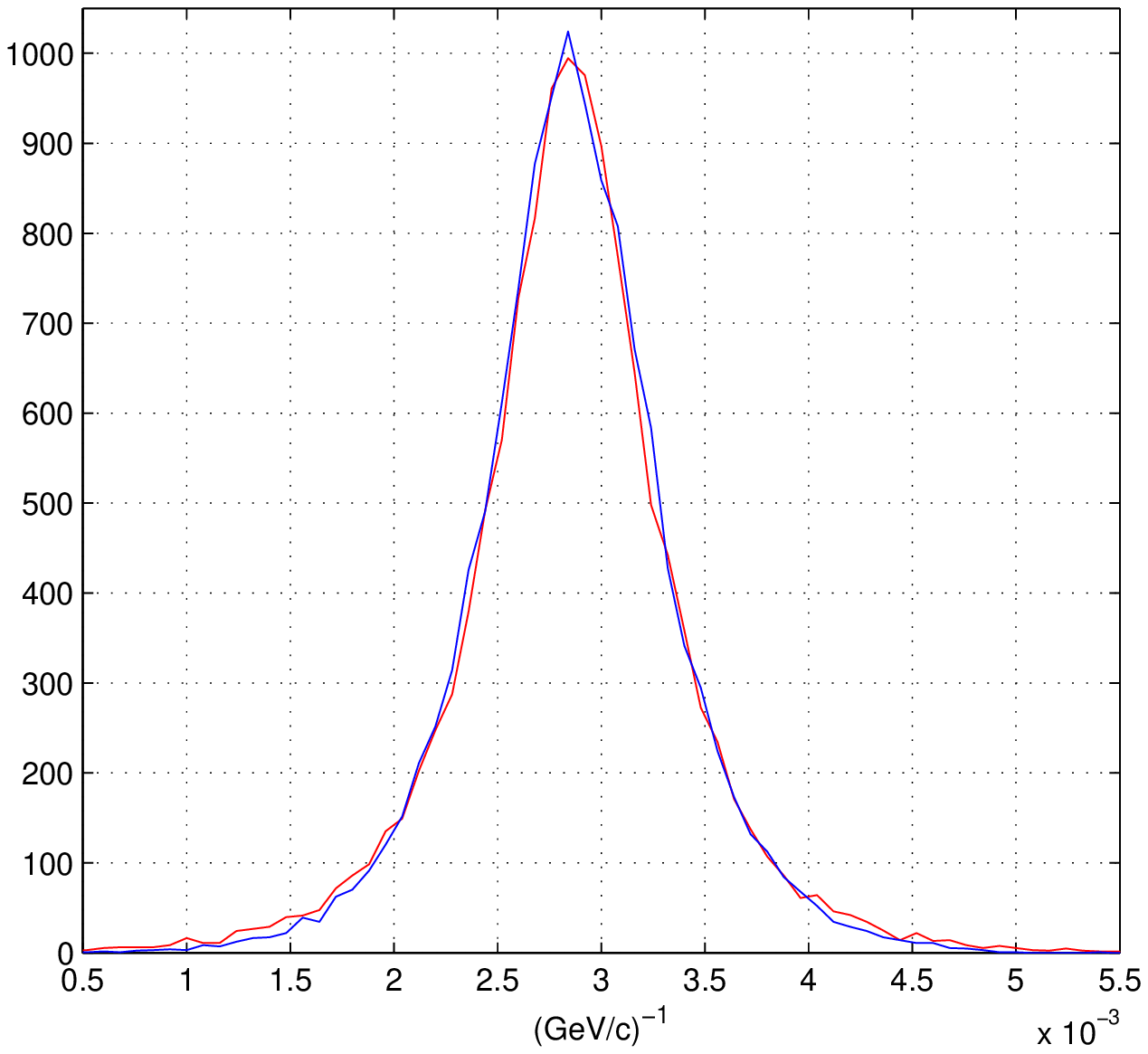}
\includegraphics[scale=0.45]{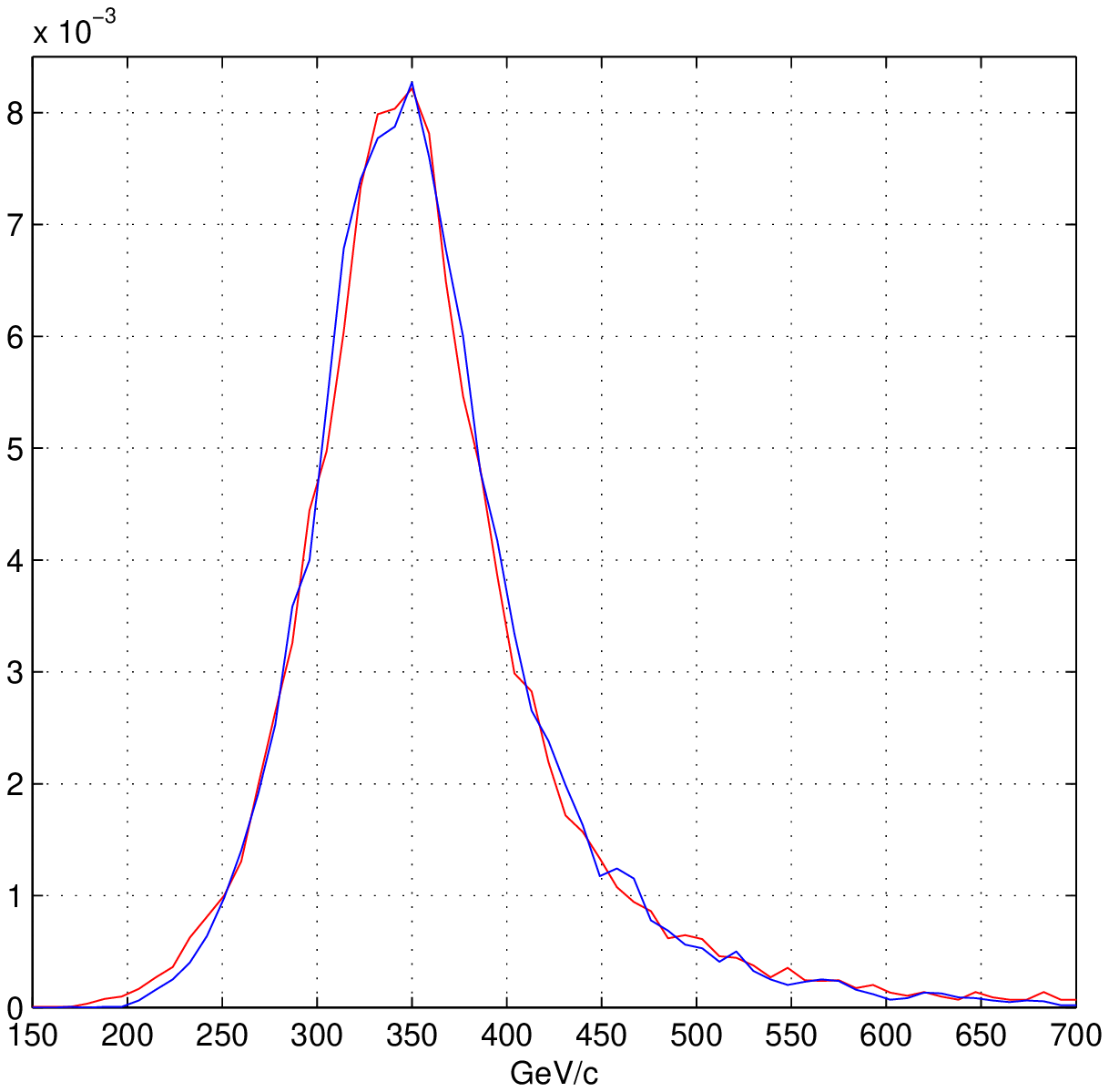}
\includegraphics[scale=0.45]{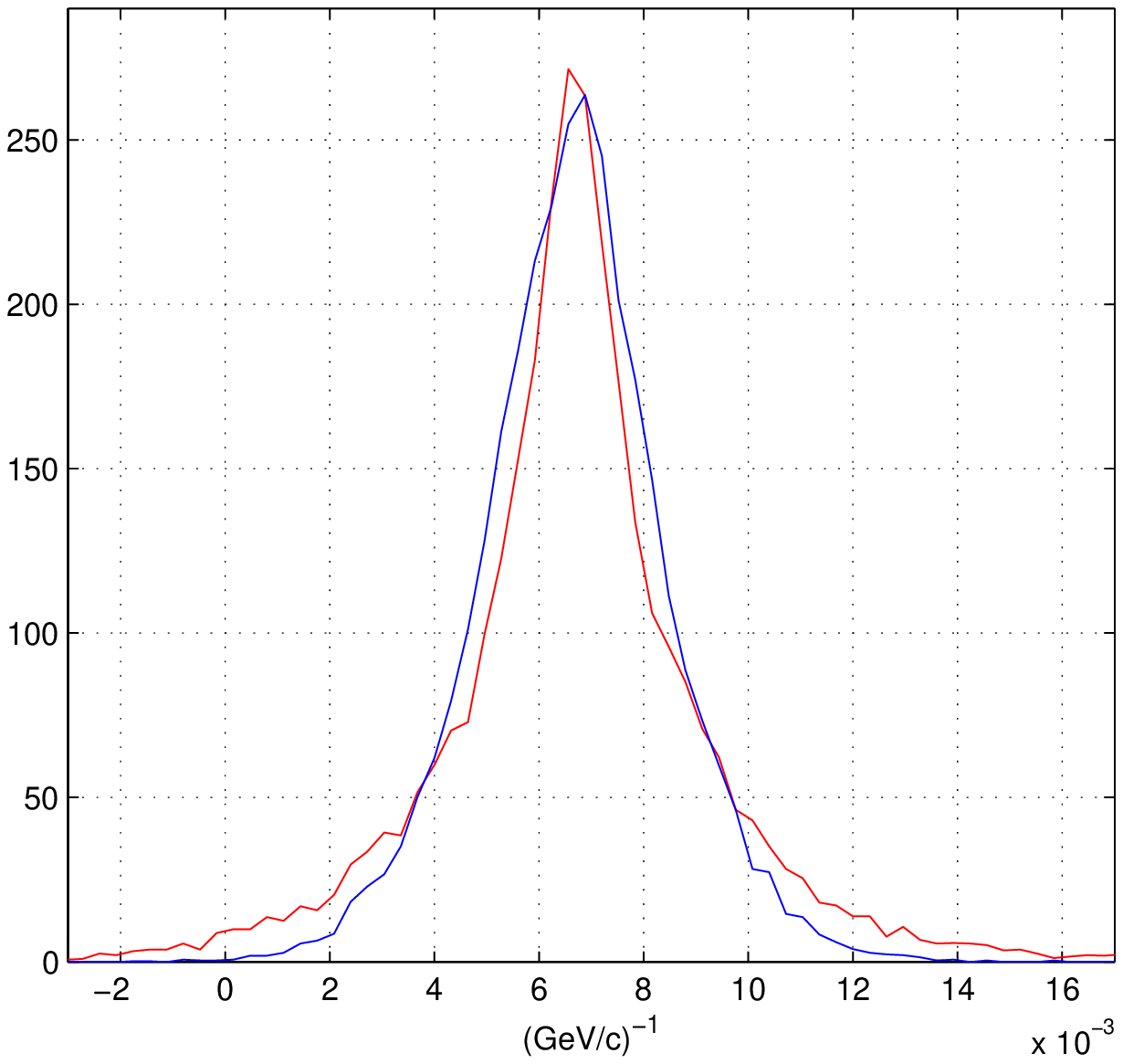}
\includegraphics[scale=0.45]{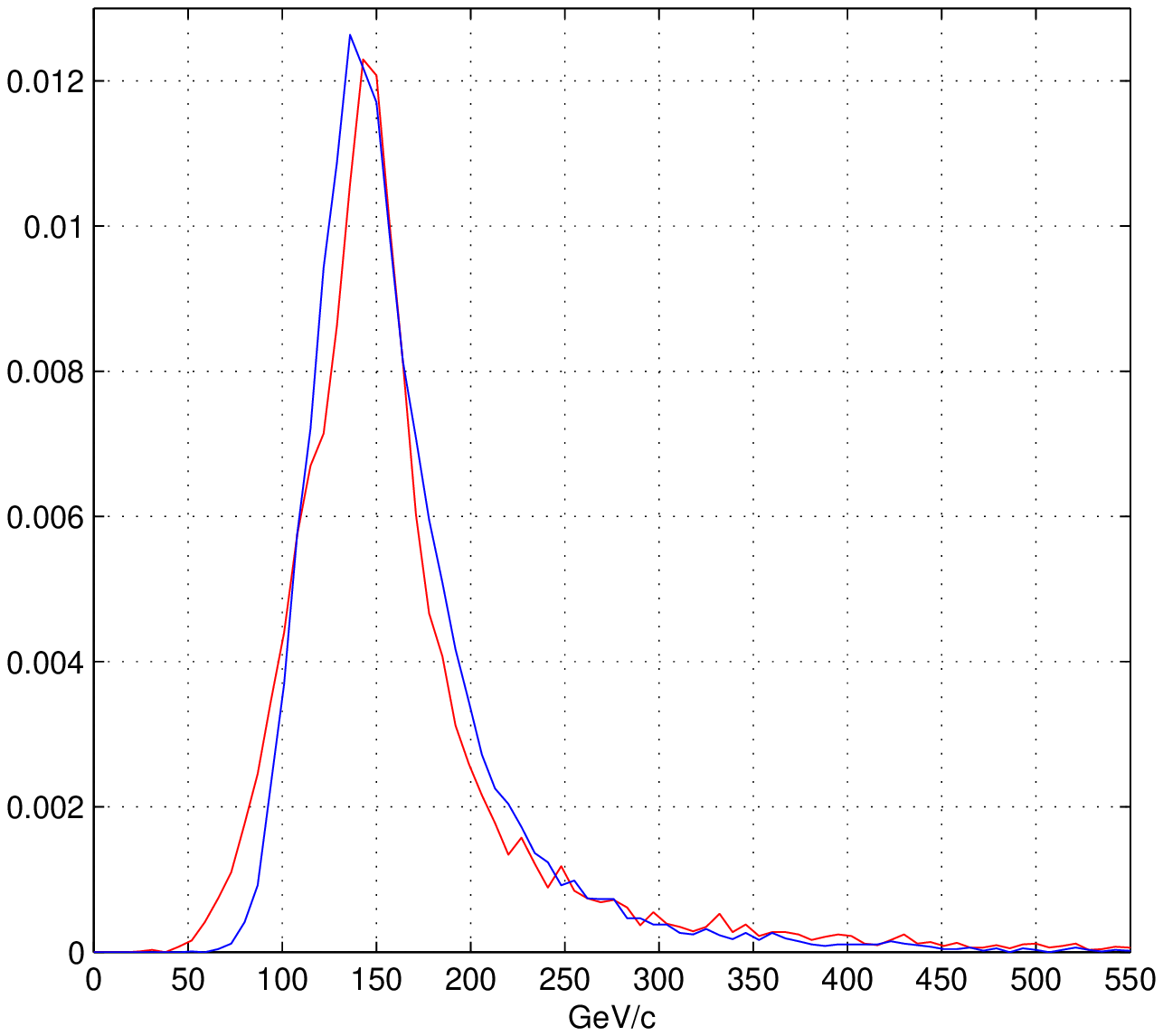}
\caption{\em Top plots. Low noise case, Curvature in $(GeV/c)^{-1}$ and momentum distributions for our MLE (red lines)
and $\eta_2$ least squares with a magnetic field increased by a factor 1.5. Bottom plots. Higher noise case, here the $\eta_2$
least square has a magnetic field 1.8 times greater than the MLE.
}\label{fig:figure05}
\end{center}
\end{figure}

With fixed momentum, the magnetic field is increased in the fit for the $\eta_2$ least squares.
The upper sector of fig.~\ref{fig:figure05} illustrates these results for the low noise side and
the overlaps with our
MLE. The red lines
are identical to those of fig.~\ref{fig:figure02}, the blue line are the
$\eta_2$ least squares for a magnetic field 1.5 times higher. The plots for the COG$_2$ least squares are not
reported to render easily legible the figures, in this case the magnetic field must be increased of  factor
1.8  to overlap our red lines.
The lowest sector of fig.~\ref{fig:figure05} reports the higher noise side, the red lines are identical to those of fig.~\ref{fig:figure04}, the blue lines ($\eta_2$ least squares) overlap the red lines with a magnetic field increased of 1.8 times.
Even here, the overlap with the COG$_2$ least squares is not reported, now, to obtain a reasonable
overlap, the magnetic field must be doubled. Clearly, the increment of the magnetic field produces other slight
differences in the tails of the distributions, but the main results are well evidenced.

\subsection{Increase of the signal-to-noise ratio}

Let us see now the effects of the increase of the signal-to-noise ratio.
These modifications reproduce in part the results
of the magnetic increase, but it is different in other parts. In fact, higher values  of
the magnetic field do not modify the form of the distributions of the curvature $\alpha$
(with dimension ${legth}^{-1}$),
the curvature distributions translate toward
higher values for a reduction of the track radius.
The dimensional transformations to $(\mathrm{GeV/c})^{-1}$ and $\mathrm{GeV/c}$ have the magnetic field intensity as scaling factor,
that can rise the distributions to the forms of fig.~\ref{fig:figure05} even for the COG$_2$ case.
But, as for the $\alpha$  parameters (with dimension ${length}^{-1}$),
the distributions of the true residuals of fig.~\ref{fig:figure01} and fig.~\ref{fig:figure03} are not
influenced by the rising of the magnetic, a part small effects on the Lorentz angle.

Keeping the magnetic field to $0.43\, T$, the raise of the signal-to-noise ratio modifies the
distributions of the $\alpha$ parameters and those of the true residuals and can bring them to overlap our red distributions.
In our simulations, this is accomplished scaling the amplitude of the random noise,
added to the strip signals, and rerunning the least squares fits. The plots of these results for the momentum (in $\mathrm{GeV/c}$)
and curvature (in $(\mathrm{GeV/c})^{-1}$) are not reported because they are practically identical to those
of fig.~\ref{fig:figure05}.
For the low noise side the  gaussian strip noise of $\sigma=4$ ADC counts must be reduced to 2.5 ADC counts,
raising the signal-to-noise ratio by a factor 1.6 to a huge most probable value of $SNR(n)=56.8$.
Now, even the blue line in fig.~\ref{fig:figure01} of the true
residuals reaches the red line, but the redundancy continues to be unable to move the true residual
distribution beyond that of the new $\eta_2$ errors.
A special mention must be devoted to the COG$_2$ distributions, the magenta lines.
They remain essentially unchanged in any plot for a (any) reduction of the strip noise. The COG$_2$ error distribution
of fig.~\ref{fig:figure01} (the green line) shows a slight modification becoming less rounded for the hard presence
of the systematic error, unaffected by the random noise.

For the higher noise side of the detector, a reduction to less of one half of the
noise (from $8$ to $3.6\ \ \mathrm{ADC}$-counts) is required to reproduce the low parts of fig.~\ref{fig:figure05}.
With this doubling (2.2) of the signal-to-noise
ratio the blue line (the $\eta_2$ best fit) overlaps the red line in fig.~\ref{fig:figure03} rising the most probable
signal-to-noise ratio to 40.5.
As for the other side, no noise reduction moves the distributions given by the COG$_2$ least-squares. The magenta lines remain practically unchanged a part a reduction of the fluctuations and a sharpening of the green line of fig.~\ref{fig:figure03}.

As just said above, the insensitiveness to the noise  reduction is due to the COG systematic error
of ref.~\cite{landi01}. This fact renders of weak relevance the efforts to
improve the signal-to-noise ratio in the detectors if the positioning algorithm
remains the COG. On the other side, this method is stable respect to
a decrease of the signal-to-noise ratio due to ageing or radiation damages,
as far as the COG systematic error dominates.

A last point remains to be analyzed, i.e. the huge difference between the curvature PDF
for the $\eta_2$ least squares in the low part fig.~\ref{fig:figure05} and the blue line of fig.~\ref{fig:figure02}.
Now the two detector types have very similar signal-to-noise ratio (3.6 ADC counts in the first and
4 ADC counts in the second), but, to reach the overlap of the two PDFs, another factor of around 2.4 is
needed. This factor is too large to be due to the $20\,\%$ of differences between
the sizes of their strips. The main difference must be due to the beneficial charge sharing of the floating strip
and the nearing of its detector architecture to the ideal detector defined in ref.~\cite{landi01}.

\section{Conclusions}

We extended the simulated track reconstructions of ref.~\cite{landi05} with the study of curved tracks in a
constant magnetic field of $0.43\, T$. All the other effects, $\delta$-rays,
multiple scattering (negligible at our high momenta), energy loss,
etc., are explicitly excluded by the simulation, focusing on the differences among various fit methods.
Each type of detectors of our double sided detector are examined as momentum analyzer.
The higher noise side has  signal properties very similar to single sided
detectors of large use in running trackers.
We use our well tuned PDFs in two form: complete in the MLE or schematic with weight parameters $1/\sigma_{eff}(i)^2$.
Each one of these two form is very effective for this task
respect to the results of two other type of track fitting explored i.e. the least squares with $\eta_2$ as position
algorithm and the least squares with the COG$_2$ positioning.
To establish a comparison among these fitting methods, we skip the usual ways of comparing standard deviations or
full width at half maximum. Instead, we modify two different tracker properties to reach
an overlap among the fit outputs: the magnetic field and the signal-to-noise ratio.
For the overlap of the two standard fits with our best distributions,
the magnetic field must be increased by a factor 1.5 for the $\eta_2$ fit and 1.8 for  the COG$_2$ in the low
noise side and 1.8 and 2 for the higher noise side.
The increase of signal-to-noise ratio is effective only for the $\eta_2$ based least squares, the overlaps are obtained
with  factors 1.6 and 2.2  for the two detector sides.
Any increase of the signal-to-noise ratio is unable to improve the curvature and momentum
distributions in the case of COG$_2$ positioning, the COG intrinsic
systematic error survives untouched to any reduction of the detector random noise.
In any case, the drastic improvement
of our well tuned PDFs in the momentum reconstructions  is evident. We have to remind that these are simulations and come
with all their
uncertainties. Assuming an optimistic view, this increase in resolution can be spent in different ways
either for better results on running experiments or in reducing the complexity of future experiments if the
baseline fits (almost always based on the COG positioning) are estimated sufficient.
With eq.~\ref{eq:equation_1c}, we start to give a glimpse of the analytical forms of our PDFs.
Even if this expression is of limited validity, its combination with the appropriate $\{a_j(\varepsilon)\}$
allows a faithful reproduction of the green and cyan distributions of figs.~\ref{fig:figure01} and~\ref{fig:figure03}
from the data. Those distributions are exclusive products of simulations, but, with slight redefinitions,
the simplified PDF is able to reproduce them. The complex task of MLE requires
more advanced expressions, future papers will be devoted to their discussion.

\end{document}